**ARTICLE** OPEN

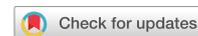

# Measurement-device-independent quantification of irreducible high-dimensional entanglement

Yu Guo[1,2,3], Bai-Chu Yu[1,2,3], Xiao-Min Hu[1,2], Bi-Heng Liu[1,2✉], Yu-Chun Wu[1,2✉], Yun-Feng Huang[1,2], Chuan-Feng Li[1,2✉] and Guang-Can Guo[1,2]

The certification of entanglement dimensionality is of great importance in characterizing quantum systems. Recently, it was pointed out that quantum correlation of high-dimensional states can be simulated with a sequence of lower-dimensional states. Such a problem may render existing characterization protocols unreliable—the observed entanglement may not be a truly high-dimensional one. Here, we introduce the notion of irreducible entanglement to capture its dimensionality that is indecomposable in terms of a sequence of lower-dimensional entangled systems. We prove this new feature can be detected in a measurement-device-independent manner with an entanglement witness protocol. To demonstrate the practicability of this technique, we experimentally apply it on a 3-dimensional bipartite state, and the result certifies the existence of irreducible (at least) 3-dimensional entanglement.

*npj Quantum Information* (2020)6:52 ; https://doi.org/10.1038/s41534-020-0282-4

## INTRODUCTION

High-dimensional (HD) quantum entanglement provides tremendous advantages over its two-level counterpart in many quantum information processes[1–4]. As a key feature in characterization of entanglement, great efforts have been made to explore the dimensionality of it[5,6], which is termed as the minimum local Hilbert space dimension required to faithfully represent its correlations.

Intuitively, the violation of a *d*-dimensional Bell-type inequality[7–9] or the surpassing of the upper bound of a *d*-dimensional entanglement measure monotone implies a dimensionality greater than *d*. Other impressive progress is marked by the construction of entanglement dimensionality witness[10–13]. However, these methods care little about the reducibility of the dimension, which is a fine-grained notion concerning whether a high-dimensional system can be reduced into a sequence of lower-dimensional ones. Consider a ququart state and two copies of qubit states, which both have dimension 4 in a perspective of Hilbert space. However, operating on two-qubit states sequentially cannot guarantee the ability of a ququart state in information processing[14,15]. As another example, it is proved that *n*-outcome quantum correlations are stronger than any nonsignaling correlations produced from selecting among $(n-1)$-outcome measurements[16,17].

Following the notion, we say that an entangled system has irreducible dimension, or irreducible entanglement in short, if and only if it can produce correlation that cannot be simulated by sequential operations on a sequence of lower-dimensional subsystems, and thus the entanglement is a truly high-dimensional one.

When the quantum states and the quantum operations are well-characterized, it is not difficult to determine the reducibility of an entangled system. However, it is not a trivial task in practical scenarios, where the experimental apparatus are not trustable. Recently, it was pointed out that, quantum correlation of HD entanglement could be simulated through sequential operations acting on copies of qubit states with the help of classical feedback[14]. This indicates that the failure of capturing the reducibility of entanglement can lead to unreliability of existing device-independent (DI) tools in detecting entanglement dimensionality, as signals used to certify HD entanglement may in fact be produced with lower-dimensional systems.

To capture this intrinsic feature of HD systems, a characterizing protocol must certify a task that cannot be accomplished with reducible systems. A solution is proposed for certifying irreducible entanglement dimension 4 (two-ququart system) in ref. [14], which is based on certifying the existence of entangled measurement between local subsystems. However, it is inapplicable for general entangled states and requires extreme quantum correlations. Also, it requires assumptions on the local subsystems and the two distant parties. Therefore, a general tool for irreducible entanglement witness is still lacked.

Here, we address these problems and show that a recently proposed protocol called quantitative measurement-device-independent entanglement witness ($Q$MDIEW)[18,19] can provide a general method to quantify irreducible entanglement. We firstly introduce the task of discriminating irreducible entanglement and then prove that in a $Q$MDIEW, the quantified entanglement exceeds the upper bound of *m*-dimensional systems if and only if the shared system is entangled in irreducible dimension at least $m+1$. Then, we experimentally demonstrate the protocol on a two-qutrit system and observe a lower bound of its generalized robustness (GR)[20] exceeds the value of arbitrary 2-dimensional systems. Our result shows the existence of irreducible HD entanglement, marking an important step beyond previous demonstrations of MDIEW on qubit systems[21–23].

## RESULTS

### Irreducible entanglement in practical scenario

Consider a practical scenario where two deceitful experimenters, Alice and Bob, share *k* entangled states of dimension $m_i (1 \leq i \leq k)$, but claim to a referee, Charlie that their systems are entangled in a higher dimension *n*. Here, we suppose $m < n$, with $m = \max_i \{m_i\}$ and even a more general case with $\prod_{i=1}^{k} m_i = n' \geq n$ is permitted. Charlie aims to penetrate this wile and to determine the genuine dimensionality of the shared entanglement. In this case, Charlie cannot trust Alice and Bob and should treat the shared quantum state $\rho_{AB}$ and measurement apparatus as black boxes, from which he requires output answers responding to his input questions.

In a canonical DI entanglement test[7–9], the inputs are measurement labels $\{x, y\}$ and the outputs are corresponding outcomes $\{a, b\}$. Charlie determines the entanglement dimensionality of the shared entanglement based only on the statistics $p(a, b|x, y)$. However, it was pointed out in ref. [14] that correlations of HD entanglement may be simulated through a sequential procedure on copies of lower-dimensional states, thus invalidate such DI protocols. The authors gave an example that 4-dimensional Collins-Gisin-Linden-Massar-Popescu (CGLMP$_4$) inequality[24] can be violated by measuring sequentially on two copies of qubit Bell pairs with the help of locally classical feedforward of the first measurement outcomes. This example is included in the Supplementary Note 1 for completeness. The sequential procedure can be described as follows: after receiving a pair of inputs $(x, y)$, Alice and Bob perform sequential measurements $\mathcal{M}_A^{(i)}(x, i^-)$ and $\mathcal{M}_B^{(i)}(y, i^-)$ with $i = 1, 2, \ldots, k$ on their local subsystems, respectively, and obtain sequences of outcomes $S_{\boldsymbol{a}} = (a^{(1)}, \ldots, a^{(k)})$ and $S_{\boldsymbol{b}} = (b^{(1)}, \ldots, b^{(k)})$. Here, $\mathcal{M}_A^{(i)}(x, i^-)$ means that the choice of the $i$th measurement setting can be dependent on input $x$ and former $i - 1$ measurement settings and outcomes $(a^{(1)}, \ldots, a^{(i-1)})$, and $\mathcal{M}_B^{(i)}(y, i^-)$ analogously. The correlations then read

$$p(S_{\boldsymbol{a}}, S_{\boldsymbol{b}}|x, y) = \mathrm{Tr}\left(\prod_{i=k}^{1} A_a^{(i)} \otimes B_b^{(i)} \rho_{AB}\right), \quad (1)$$

where $\rho_{AB} = \rho_{A_1 B_1} \otimes \ldots \otimes \rho_{A_k B_k}$ is the composite state of $k$ entangled systems, $A_a^{(i)}$ and $B_b^{(i)}$ are the measurement operators corresponding to outcome $a_i, b_i$ of measurement $\mathcal{M}_A^{(i)}(x, i^-)$ and $\mathcal{M}_B^{(i)}(y, i^-)$, performed nontrivially (not always identities) on state $\rho_{A_i B_i}$ and being identities on states $\rho_{A_j B_j} (j \neq i)$. Then the experimenters report the outcomes to Charlie after applying some post-processing function. Since it is assumed that Alice and Bob cannot communicate during the experiment, the post-processing function should be written independently as $p(a|S_{\boldsymbol{a}})$ and $p(b|S_{\boldsymbol{b}})$. Then the form of correlations that Alice and Bob finally produce is

$$p(a, b|x, y)_m^n = p(a|S_{\boldsymbol{a}}) p(b|S_{\boldsymbol{b}}) p(S_{\boldsymbol{a}}, S_{\boldsymbol{b}}|x, y), \ \forall a, b, x, y, \quad (2)$$

where $p(a, b|x, y)_m^n$ indicates that the correlations are produced in an experiment where a sequence of systems entangled in dimension at most $m$ are used to simulate a $n$-dimensional entangled system.

To determine the dimensionality of the shared entanglement, Charlie usually defines a function, notated as $\mathcal{F}$, on his received correlations $\{p(a, b|x, y)\}$. Generally, the function can be chosen to be Bell-type parameters, witness observables, and entanglement measures. Under such situation, Charlie can confirm that the entangled system has irreducible dimension $n$ if the performance of correlations $\{p(a, b|x, y)\}$ under the function $\mathcal{F}$ is better than the correlations producible by any lower-dimensional entangled systems, which can also be written as

$$\mathcal{F}(\{p(a, b|x, y)\}) > \overline{\mathcal{F}}(\{p(a, b|x, y)_m^n\}), \ \forall m < n \quad (3)$$

where $\overline{\mathcal{F}}$ means the upper bound. Since it is shown that the correlation produced in a DI entanglement test cannot guarantee the relation in Eq. (3), Charlie cannot distinguish whether the entanglement is reducible. In ref. [14], a solution is proposed to certify that a entangled system of dimension 4 is irreducible into dimension 2, by requiring one local party (say Bob) to certify an entangled measurements between the local qubit subsystems, which cannot be accomplished by sequential operations on them. However, its robustness relies on assumptions on the local subsystems, namely, it must constrain the form of post-processing function $p(a|S_{\boldsymbol{a}})$ and $p(b|S_{\boldsymbol{b}})$. Since in the scenario the experimenters are considered not trustable, we should make improvements to decrease the needed assumptions on them.

As the unreliability of DI protocols in detecting irreducible entanglement mainly stems from the classical inputs, which are perfectly distinguishable and only classically correlated to the measurement operators, we consider replacing them with quantum inputs as that has been done in the measurement-device-independent (MDI) protocols[25]. Here, the quantum inputs are chosen from some set of nonorthogonal quantum states, which are not perfectly distinguishable, and interact to the measurement operators in quantum way. In the following, we show that in a $\mathcal{Q}$MDIEW, the quantified entanglement exceeds the upper bound of $m$-dimensional systems if the shared system is entangled in irreducible dimension at least $m + 1$, even if the post-processing function is taken as a much more general form written as $p(a, b|S_{\boldsymbol{a}}, S_{\boldsymbol{b}})$ and without further constraints. In return, the cost is that Charlie must prepare trusted input states of genuine dimension $n$ and Alice and Bob must perform entangled measurement of dimension $n$ between the input system and their own systems.

## Framework of $\mathcal{Q}$MDIEW

We briefly introduce the construction of the $\mathcal{Q}$MDIEW, defined as a MDIEW whose expectation on a state provides a lower bound on its entanglement. Here, the blackbox-like devices receive pairs of quantum states $\tau_x, \tau_y$ chosen randomly from a given set $\{\tau_1 \ldots \tau_n\}$. To certify the shared entanglement, Alice (Bob) performs a joint measurement described by some positive-operator-valued measurement (POVM) $\{M_a\}$ or $\{M_b\}$ on her (his) own system and input states. Correlations in this case can be written as

$$p(a, b|\tau_x, \tau_y) = \mathrm{Tr}[(M_a \otimes M_b)(\tau_x \otimes \rho_{AB} \otimes \tau_y)], \quad (4)$$

where $a$ ($b$) is the outcome of $M_a$ ($M_b$). The system above can also be described with an effective POVM $\{\Pi_{ab}\}$ acting on input states $\tau_x \otimes \tau_y$. In this case, correlations in Eq. (4) can be generally expressed as

$$p(a, b|\tau_x, \tau_y) = \mathrm{Tr}[\Pi_{ab}(\tau_x \otimes \tau_y)]. \quad (5)$$

It is proved that the POVM $\{\Pi_{ab}\}$ can be used to recover the ensemble, which can be extracted from the setup and whose entanglement provides a lower bound on the entanglement of the shared state $\rho_{AB}$ (ref. [19]). That is, the entanglement of $\rho_{AB}$ under an arbitrary convex entanglement $\mathcal{E}$ (ref. [26]) is lower bounded by applying the measure $\mathcal{E}$ on the set of operators $\{\Pi_{ab}\}$, formulated as

$$Q = \frac{1}{d_X d_Y} \sum_{ab} \mathcal{E}(\Pi_{ab}) \leq \mathcal{E}(\rho_{AB}), \quad (6)$$

where $d_X$ ($d_Y$) is the dimension of the Hilbert space $\mathcal{H}_X(\mathcal{H}_Y)$ of Alice's (Bob's) input state and $Q$ is the lower bound obtained in this method. The entire process from raw correlations in Eq. (5) to a quantified entanglement $Q$ (a lower bound) of $\rho_{AB}$ can be described as a semidefinite programming (SDP[27]) and a $\mathcal{Q}$MDIEW can be recovered from its dual process.

## $\mathcal{Q}$MDIEW captures irreducible entanglement

We now consider the correlations Alice and Bob can produce through the sequential procedure in $\mathcal{Q}$MDIEW. In this protocol, to examine whether the experimenters have a entangled system of irreducible dimension $n$, Charlie prepares the quantum inputs $\{\tau_x, \tau_y\}$, which are single-particle $n$-dimensional quantum states. As Alice and Bob cannot directly get information about the input, the sequence of measurement setting can be written as $\mathcal{M}_{XA}^{(i)}(i^-)$ and $\mathcal{M}_{YB}^{(i)}(i^-)$, and as they cannot create entanglement using local operations and classical communication[28], the post-processing function can be written as $p(a, b|S_a, S_b)$, and the correlations they can produce are

$$p(a, b|\tau_x, \tau_y)_m^n = \sum_{S_a, S_b} p(a, b|S_a, S_b) p(S_a, S_b|\tau_x, \tau_y), \tag{7}$$

with

$$p(S_a, S_b|\tau_x, \tau_y) = \text{Tr}\left(\prod_{i=k}^{1} M_a^{(i)} M_b^{(i)} \tau_x \otimes \rho_{AB} \otimes \tau_y\right), \tag{8}$$

where $M_a^{(i)}$ and $M_b^{(i)}$ are the measurement operators of $\mathcal{M}_{XA}^{(i)}(i^-)$ and $\mathcal{M}_{YB}^{(i)}(i^-)$, performed nontrivially on the input systems and their $i$th entangled systems, and being identities on the other entangled systems. Since Alice and Bob are simulating a high-dimensional bipartite system, we represent composite system of $k$ subsystems as one higher-dimensional bipartite system, and represent the composite quantum state and measurement operators in the basis of this higher-dimensional system. Let $O^d$ denotes the space of operators on $d$-dimensional Hilbert space, then the correlations in Eq. (8) can be written into an equivalent form as

$$p(S_a, S_b|\tau_x, \tau_y) = \text{Tr}[(\overline{A}_{S_a}^m \otimes \overline{B}_{S_b}^m)(\tau_x \otimes \overline{\rho}_{AB} \otimes \tau_y)], \tag{9}$$

where $\overline{\rho}_{AB} \in O^{n'} \otimes O^{n'}$ is the corresponding state of $\rho_{AB}$, $\overline{A}_{S_a}^m \in O^{n'} \otimes O^{n'}$ are the corresponding measurement operator of $\prod_{i=k}^{1} M_a^{(i)}$ and $\overline{B}_{S_b}^m$ in analogy. In QMDIEW protocol, when Charlie obtains the correlation $p(a, b|\tau_x, \tau_y)_m^n$ produced by Eqs. (7) and (8), he reconstructs the operators $\Pi_{ab}$ and computes the entanglement using Eqs. (5) and (6). We can see that the computed entanglement depends on the entanglement of $\Pi_{ab}$, and the entanglement of $\Pi_{ab}$ depends on $\overline{\rho}_{AB}$, $\overline{A}_{S_a}^m$ and $\overline{B}_{S_b}^m$. Now, we prove that the entanglement Charlie finally detected in QMDIEW cannot exceed the upper bound of $m$-dimensional entanglement, which is denoted by $\mathcal{E}^m$. The demonstration includes two theorems, whose proves are shown in Supplementary Information.

**Theorem 1** When Alice and Bob perform the sequential procedure on entangled systems of dimension at most $m$ in a QMDIEW protocol, the measurement operators $\overline{A}_{S_a}^m$ and $\overline{B}_{S_b}^m$ satisfy $\mathcal{E}(A_{S_a}^m) \leq \text{Tr}(A_{S_a}^m) \cdot \mathcal{E}^m$, $\mathcal{E}(B_{S_b}^m) \leq \text{Tr}(B_{S_b}^m) \cdot \mathcal{E}^m$.

Theorem 1 indicates that in a QMDIEW, the measurement operators Alice and Bob can simulate is strongly restricted with a bounded entanglement. In fact, we can see from the proof (see Supplementary Note 2) that the entanglement of the effective operators depends only on the first joint measurement of Alice and Bob, the rest of the measurements and communications will not increase this entanglement.

**Theorem 2** In the above QMDIEW game, the detected entanglement will never exceed the upper bound of $m$-dimensional entanglement, i.e., $\mathcal{E}_{MDI}(\{p(a, b|\tau_x, \tau_y)_m^n\}) \leq \mathcal{E}^m$ when the quantum input set $\{\tau_x, \tau_y\}$ is tomographically complete.

Theorem 2 (see Supplementary Note 3 for the proof) gives an important result that if the entanglement Charlie obtained exceeds the bound of $m$-dimensional bipartite states, he can confirm that Alice and Bob share an entanglement with irreducible dimensionality of at least $m + 1$. A simple explanation for this result is: to obtain a quantity exceeding the bound of $m$-dimensional entangled system, the experimenters must make entangled measurements between a $m + 1$ dimensional local system and each received input state. Such measurement cannot be completed sequentially, since an entangled measurement collapses the input state. Now, we conclude that QMDIEW not only certifies the quantity of the shared entanglement but also captures its irreducible dimension.

Although assuming that the input states are trustable as in standard MDI tests[25,29], our method here is more experiment friendly than the present method in ref. [14], where an additional selftest process was introduced to certify an entangled measurement. Firstly, our method do not require extreme quantum correlations as selftest technique requires. Secondly, MDI protocols does not suffer to locality loophole in practical scenario. Moreover, our method removes the need of any constraints on the post-processing function, which means that we have greatly reduced the required additional assumptions on the two distant systems and between local subsystems. The reason that such post-processing function is allowed is that the SDP includes a minimization among all possible POVM operators $\Pi_{ab}$ to calculate entanglement from correlations $p(a, b|\tau_x, \tau_y)$. Therefore, any post-processing functions just serve as coefficients of some combination of the POVM operators and does not increase the entanglement of the unprocessed correlations. See Supplementary Note 4 for a detailed illustration.

### Experimental protocol

We consider the experimental observation of irreducible HD entanglement with a simplest system composed of a two-qutrit state. To simulate 3-dimensional entanglement states, Alice and Bob should adopt the sequential strategy with qubit pairs. In the QMDIEW game, however, Charlie can identify the system is truly entangled in dimension 3 as long as he observes its entanglement exceeding the upper bound of two-qubit systems. Here, the entanglement measure is chosen to be GR, whose value is upper bounded by $n - 1$ for bipartite $n$-dimensional systems[20].

In our experiment, the state is chosen to be a 3-dimensional maximally entangled state, i.e., $\rho_{AB} = |\phi\rangle\langle\phi|$, with $|\phi\rangle = \frac{1}{\sqrt{3}}(|00\rangle + |11\rangle + |22\rangle)$. After the distribution of $\rho_{AB}$, Charlie prepares input states $\tau_x$ and $\tau_y$ randomly selected from a set $S \equiv \{|0\rangle, |1\rangle, |0\rangle + |1\rangle, |0\rangle + i|1\rangle, |0\rangle + |2\rangle, |1\rangle + |2\rangle\}$ (this input set (set $S$) does not generate tomographically complete quantum inputs for the shared two-qutrit state, but by calculation we found that it suffices to produce the same entanglement quantity under full Bell state measurement (BSM) for both sides (81 outcomes). So we choose input states from set $S$ to lessen the complexity of experiment) and sends them to Alice and Bob. Then Alice and Bob choose to perform POVM $\{A_a\}$ and $\{B_b\}$ consisting of the projectors onto all nine Bell states on the distributed qutrit and input state. The obtained correlations $p(a, b|\tau_x, \tau_y)$ are reported to Charlie who then performs a regularization process to find out the closest regularized distribution $p_r(ab|\tau_x, \tau_y)$ and thus corrects the inconsistencies of raw $p(ab|\tau_x, \tau_y)$ caused by noise and finite statistics[21]. Based on $p_r(ab|\tau_x, \tau_y)$, Charlie can calculate the GR of $\rho_{AB}$ and recover a QMDIEW (see Supplementary Notes 5 and 6). With the quantified GR of $\rho_{AB}$, Charlie can finally determine a lower bound on its irreducible dimension of entanglement according to Theorem 2.

We realize this procedure using linear optics. The whole-experimental setup is illustrated in Fig. 1 and can be divided into three modules: the Source module for state preparation (orange), the Trust module for trusted inputs preparation (blue) and the BSM for 3-dimensional BSM (green). Experimental details are presented in the "Methods" section.

### Results analysis

We calculate the entanglement of $d$-dimensional isotropic Bell states, i.e., $\rho_{iso}^d = \frac{p}{\sqrt{d}}\sum_{i=0}^{d-1}|ii\rangle\langle ii| + (1-p)I/d$. Figure 2 shows the results of quantified entanglement through the $\mathcal{Q}$MDIEW protocol, where GR as an entanglement measure varies with different Bell state fraction $p$. The calculated GR of 3- and 2-dimensional noisy Bell states are given in the red and blue lines, respectively, and match well with their theoretical predictions. This implies all entangled states can be detected in a MDI way through the protocol. When $p = 1$, the values of GR reach maximums for both 3- and 2-dimensional cases and are 2 and 1, respectively. Our experimental result of the state $\rho_{AB}$ is marked with the black dot and the obtained GR is $1.632 \pm 0.017$. The deviation of the GR value from its theoretical predictions are due to imperfect state preparation and measurement error. The experimental fidelity of $\rho_{AB}$ is $f_{real} = 0.986 \pm 0.002$ (here the fidelity is estimated by the presented data, and we have not done a complete state tomography).

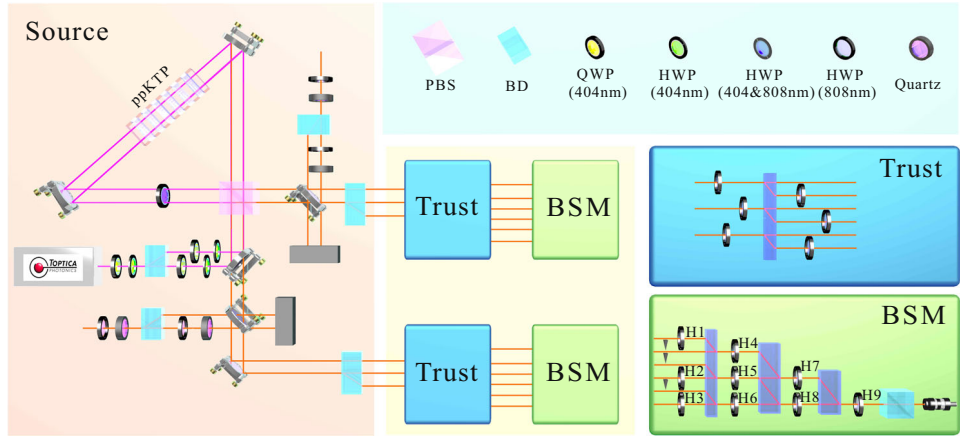

**Fig. 1 Experimental setup.** The setup contains three modules: the Source module, the Trust module, and the BSM module. In the Source module, 3-dimensional maximally entangled state $|\phi\rangle = \frac{1}{\sqrt{3}}(|00\rangle + |11\rangle + |22\rangle)$ and maximal mixed state $I/9$ are prepared, respectively, and the two states are then combined with different proportions at the beam splitters. The photon pair rate was around 2500 counts/s with a detection efficiency of 0.222. In the Trust module, arbitrary 3-dimensional pure state can be prepared by expanding the distributed photons' path degree of freedom. High quality of the input state is ensured by highly controllable HWPs and BDs. In the BSM module, projector onto arbitrary one of the nine 3-dimensional Bell states can be constructed by setting angles of HWPs and blocking proper pathes. BD-beam displacer, PBS-polarizing beam splitter, HWP-half wave plate, QWP-quarter wave plate.

Based on the quantified GR, we then determine the irreducible dimensionality of the shared entanglement. As show in Fig. 2, states located in areas with different colors mean different irreducible dimensionality $d$—red represents $d = 2$, green represents $d = 3$, and blue represents $d = 4$. Theoretically, GR of the state $\rho_{iso}^d$ with $d = 3$ and $p > 5/8$ exceeds the upper bound of two-qubit systems (dash line). In our experiment, the observed GR of $\rho_{AB}$ exceeds the bound of arbitrary 2-dimensional system by >37 standard deviations and thus confirms that our system is truly entangled at least in dimension $d = 3$, which means it cannot be simulated with qubit systems through the sequential procedure. In other words, we demonstrate an irreducible 3-dimensional entangled state without characterizations on the measurement apparatus.

## DISCUSSION

In this article, we displayed the power of the $\mathcal{Q}$MDIEW protocol for HD systems by showing that it can witness irreducible entanglement, which cannot be simulated by lower-dimensional systems through the sequential procedure. Our results shows that $\mathcal{Q}$MDIEW not only can be used to quantify how much entanglement there is, but also can help to determine the genuine dimension that the entanglement exists in. As the simplest application, we demonstrated this protocol with a 3-level bipartite entangled state and quantified a lower bound of its GR that exceeded the bound of 2-dimensional systems, presenting the existence of irreducible 3-dimensional entanglement.

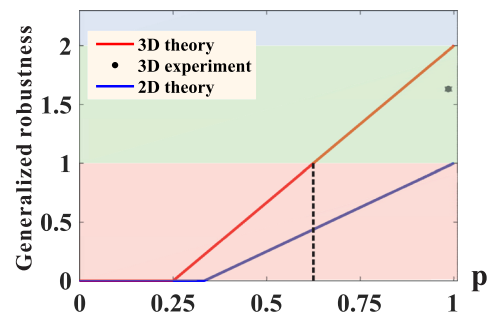

**Fig. 2 GR varies as a function of the weights $p$.** The red and blue lines are calculated theoretical results for 3- and 2-dimensional isotropic Bell states $\rho_{iso}^d$. For 3-dimensional case, GR rises linearly from 0 to 2 when $p$ varies from 1/4 to 1, while it rises from 0 to 1 with $p$ varies from 1/3 to 1 for 2-dimensional case. The black dot is our experimental result which is $1.632 \pm 0.017$. Error bar is simulated by Monte Carlo algorithm. Different color of background indicates different irreducible dimension $d$ of the shared entanglement. Red means $d = 2$, green means $d = 3$, and blue means $d = 4$.

Irreducible entanglement here differs from the notion of genuine HD entanglement proposed in ref. [30] as the later removes the sequentiality assumption in definition. Genuine HD entanglement is more stringent as it rules out low-dimensional states's and operations' responds for observed behavior independently of the simulation strategies. As a simple example, a 4-dimensional maximally entangled state is irreducible 4-dimensional while it is not genuine 4-level entanglement as it is decomposable with two copies of Bell states.

Our work may provide inspiration for future works. From a theoretical point of view, we see that in a $\mathcal{Q}$MDIEW, Alice and Bob have to perform entangled measurements of dimension $d$ to certify the irreducible dimension $d$, which is an extra cost. In return, the need of trust on them can be removed. It may be an interesting task to study the tradeoff relation between the cost of entangled measurement and the trust on it. From a practical point of view, it is of great importance to realize unconditional MDI demonstrations. Although the MDI method can tolerate low detection efficiencies, it would in turn reduce the observed entanglement. This could be strengthened by highly efficient superconducting detectors[31] and a complete high-dimensional Bell state measurement. Also, our experiment leads to important applications based on high-dimensional entangled systems, such as randomness generation[32,33] and high-dimensional quantum key distribution[34], without assumptions on the measurement apparatus.

## METHODS

### Experimental details

The experimental setup is shown in Fig. 1. A cw violet laser at 404 nm was separated with a beam displacer (BD40), and then was incident to a Sagnac interferometer to pump a type-II cut ppKTP crystal to generate photon pairs at 808 nm. Through adjustments of half- and quarter-wave plates (HWPs and QWPs), 3-dimensional pure state $|\phi\rangle = \frac{1}{\sqrt{3}}(|HH\rangle_u + |HH\rangle_l + |VV\rangle_l)$, i.e., $\rho_{AB}$, was prepared, where the subscripts $u$ and $l$ denote upper path and lower path (refer to our previous works refs. [4,35] for more details). To prepare the mixed state $\rho_{noise}$, another photon pair was used[21]. Setting the angles of HWPs in the Source module, the state $\frac{1}{3}(|H\rangle_u + |H\rangle_l + |V\rangle_l) \otimes (|H\rangle_u + |H\rangle_l + |V\rangle_l)$ was obtained. Inserting four quartz, we destroyed the coherence of this state and obtained the mixed state $I/9$. Mixing the two states $\rho_{AB}$ and $I/9$ at beam splitters, we finally prepared the path-polarization hybrid entangled state $\rho_{noise}$ with tunable $p$ by adjusting the ratio of the two parts.

To prepare the input states $\{\tau_x, \tau_y\}$, Charlie expanded the dimension of photon's path degree of freedom by separating the photons via HWP and BD20 (half the length of BD40) in the Trust modules as shown in Fig. 1. Encoded as $|h\rangle_u, |h\rangle_l$, and $|v\rangle_l$, arbitrary trusted input states of the form $\beta_1 |h\rangle_u + \beta_2 |h\rangle_l + \beta_3 |v\rangle_l$ could be prepared by adjusting the angles of the HWPs. The input states $\tau_x$ ($\tau_y$) are assumed to be trustable and thus should be prepared with high accuracy. This could be met for our scheme restricts in highly controllable operations of polarization elements (HWPs and BDs). Our method to prepare the input states consumes no extra degree of freedom of the photons, which is indeed precious resource.

There are nine Bell states for a 3-dimensional system, which can be divided into three categories (each contains three of them) as – Category 1:

$$|\Phi_1\rangle = \frac{1}{\sqrt{3}}(|00\rangle + e^{i\varphi_0}|11\rangle + e^{i\varphi_1}|22\rangle), \quad (10)$$

Category 2:

$$|\Phi_2\rangle = \frac{1}{\sqrt{3}}(|01\rangle + e^{i\varphi_0}|12\rangle + e^{i\varphi_1}|20\rangle), \quad (11)$$

Category 3:

$$|\Phi_3\rangle = \frac{1}{\sqrt{3}}(|02\rangle + e^{i\varphi_0}|10\rangle + e^{i\varphi_1}|21\rangle), \quad (12)$$

where $(\varphi_0, \varphi_1) \in \{(0, 0), (2\pi/3, 4\pi/3), (4\pi/3, 8\pi/3)\}$ for each category.

It remains a nodus to implement high-dimensional BSM and seems unattainable without consuming auxiliary entangled resource, especially when two particles involved. Benefitting from QMDIEW's robustness against locality loophole, we overcome the problem by providing a method, which could be used to implement high-dimensional BSM performed on two subspaces of a single photon (the distributed qutrit subspace and the trusted input state subspace in our case). The BSM modules in Fig. 1 can be used to perform BSM onto arbitrary one of nine Bell states of three dimension. To implement these BSMs, photons with proper path-polarization modes were recombined step by step. For instance, blocking the second, third and fifth arms and setting the angles of HWP1-9 at 0°, 45°, 0°,0°, 45°, 0°, 0°, 22. 5°, and 27.37°, the measurement onto $\frac{1}{\sqrt{3}}(|Hh\rangle_u + |Hh\rangle_l + |Vv\rangle_l)$ (one of the states in category 1) was obtained. For other two states in category 1, it could be achieved by inserting a $\lambda/3$, a $2\lambda/3$ wave plate or a $\lambda/3$, a $\lambda/3$ wave plate at 0° in front of H8 and H9, respectively. Other BSMs can be achieved analogously. Recording the twofold coincidences between the single photon detectors of Alice and Bob, the correlations in Eq. (2) of the main text were then estimated under an iid constant production rate assumption.

## DATA AVAILABILITY

## CODE AVAILABILITY

## ACKNOWLEDGEMENTS
We thank Y.-K. Wang, D. Rosset, and X.-H. Li for valuable discussions. This work was supported by the National Key Research and Development Program of China (No. 2017YFA0304100, No. 2016YFA0301300 and No. 2016YFA0301700), NSFC (Nos. 11774335, 11734015, 11874345, 11821404, 11904357), the Key Research Program of Frontier Sciences, CAS (No. QYZDY-SSW-SLH003), Science Foundation of the CAS (ZDRW-XH-2019-1), the Fundamental Research Funds for the Central Universities, and Anhui Initiative in Quantum Information Technologies (Nos. AHY020100, AHY060300).

## AUTHOR CONTRIBUTIONS
Y.G. and B.-C.Y. contributed equally to the preparation of this article. B.-C.Y. developed the theoretical approach with assistance from Y.G. and Y.-C.W.; Y.G. performed the experiment with assistance from B.-H.L., X.-M.H., and Y.-F.H.; all of the authors analyzed the data and discussed the contents; B.-C.Y, Y.G., B.-H.L., and Y.-C.W wrote the paper with input from all authors; C.-F.L. and G.-C.G. supervised the project.

## COMPETING INTERESTS
The authors declare no competing interests.

## ADDITIONAL INFORMATION
**Supplementary information** is available for this paper at https://doi.org/10.1038/s41534-020-0282-4.

**Correspondence** and requests for materials should be addressed to B.-H.L., Y.-C.W. or C.-F.L.

**Reprints and permission information** is available at http://www.nature.com/reprints

**Publisher's note** Springer Nature remains neutral with regard to jurisdictional claims in published maps and institutional affiliations.